\documentstyle[multicol,prl,aps]{revtex}
\input epsf
\input{epsf.sty}

\begin{document}

\title{Surface barrier and bulk pinning in MgB$_2$ superconductor}
\author{M. Pissas, E. Moraitakis, D. Stamopoulos, G. Papavassiliou, V. Psycharis and
S. Koutandos}
\address{Institute of Materials Science, NCSR Demokritos,
153 10 Aghia Paraskevi Attiki Greece.}
\date{\today }
\maketitle

\begin{abstract}
We present a modified method of preparation of the new superconductor
MgB$_2$. The polycrystalline samples were characterized using
x-ray and magnetic measurements. The surface barriers control the isothermal magnetization loops in powder samples.
In bulk as prepared samples we always observed symmetric magnetization loops indicative of the presence of
 a bulk pinning mechanism. Magnetic relaxation measurements in the bulk sample
reveal a crossover of surface barrier to bulk pinning. (PACS numbers 74.60.Ge, 74.60.Jg,74.60.-w,74.62.Bf)
\end{abstract}

\begin{multicols}{2}

\section{Introduction}

Recently Akimitsu and colleagues\cite{akimitsu,nagamatsu} discovered
superconductivity with remarkably high superconducting transition temperature %
$T_{c}\sim 39$ K in the binary intermetallic magnesium boride
(MgB$_{2}$). MgB$_{2}$ crystallizes in the hexagonal
AlB$_{2}$-type structure, (space group $P6/mmm$) consisting of
alternating hexagonal layers of Mg atoms and graphite-like layers
of B atoms. Wang et al. \cite{0103181} using specific and
magnetization measurements cocluded that MgB$_{2}$ is a type II
superconductor with $H_{c2}(0)\cong 140$ kOe, small condensation
energy \
(in comparison with Nb$_{3}$Sn and YBa$_{2}$Cu$_{3}$O$_{7}$ ), $\xi _{{\rm o}%
}\cong 49$\AA , $\lambda _{{\rm o}}\cong 1850$\AA\ and a critical field $%
H_{c}\cong 2.6$ kOe.
Recent reports\cite{0102167,budko,bugoslavsky01a,cunningham01,eom01,glowacki01,mavoori01,kambara101,joshi01,0102338,larbalestier01}
on bulk MgB$_2$ samples revealed
high intergranular critical current densities and large bulk magnetic flux
pinning. Further the reduced weak-link nature
of the grain boundaries in polycrystalline MgB$_2$
underlines the applications potential of this material.
However, few works\cite{Lima01,0105271,0105545,Simon01,0106577,Patnaik,0107511}
presented convincing evidence that the MgB$_2$
is an anisotropic superconductor with an anisotropy parameter
$\gamma=(B_{{\rm c2}}^{{\rm ab}}/B_{{\rm c2}}^c)$ taking values into interval $2\leq\gamma \leq 6$, making difficulties
in the direct use of unoriented polycrystalline MgB$_2$ in the applications.
The detailed study of the vortex matter properties for this new
superconductor is very important from fundamental, as well as from
technological point of view. In the present paper, we study the MgB$_{2}$
superconductor using x-ray powder diffraction and magnetic measurements. In
our work we observed that in powder samples, surface barriers
control the magnetic irreversibility due to asymmetric magnetic hysteresis
loops. In as prepared bulk samples we observed symmetric magnetization loops
a fact which implies that the grain (or crystallites) boundaries contribute
significantly to the pinning of the flux lines.
\section{Experimental details}
MgB$_{2}$ samples were prepared by liquid-vapor to solid reaction in an
alumina crucible placed inside an evacuated, sealed silica tube. First, we
mixed thoroughly high purity Mg and B powders, with a slight excess of Mg in
order to balance the amount of Mg that is oxidized or freeze in the crucible
and the silica walls. Since the melting point of Mg is $610^{\mbox{o}}$ we
heated the sample with a rate of $\sim 10^{\mbox{o}}$ C/min up to the
melting point of Mg. We continued the heating up to 910$^{\mbox{o}}$C with a
lower heating rate ($\sim 1^{\mbox{o}}$ C/min). At 910$^{\mbox{o}}$C the
sample annealed for two hours and then we turned off the furnace.
\begin{figure}[tbp] 
\centerline{\epsfxsize=8.6cm \epsfclipon \epsffile{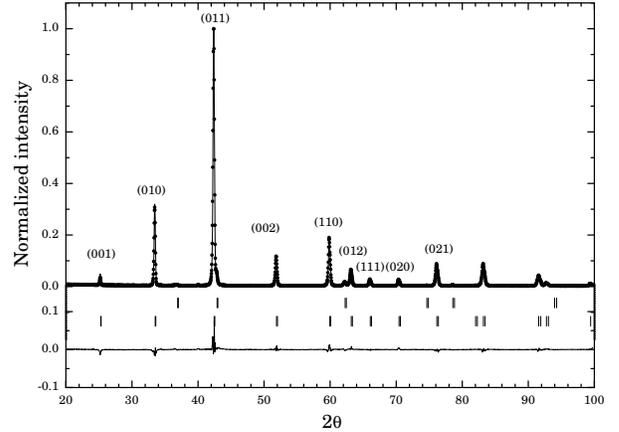}}
\caption{
Rietveld refinement pattern for the MgB$_{2}$ sample. The observed data
points are indicated with filled circles,
while the calculated pattern is shown as a continuous line. The
positions of the reflections are indicated with vertical lines below the pattern.}
\label{rietveld}%
\end{figure}%
X-ray powder diffraction (XRD) data were collected with a D500 SIEMENS
diffractometer, using CuK$\alpha $ radiation. DC
magnetization measurements were performed in a superconducting quantum
interference device (SQUID) magnetometer (Quantum Design). Ac-susceptibility
measurements were performed at zero dc field by means of a home-made probe
at a frequency of 77.7 Hz.
\section{Results and discussion}
\begin{figure}[tbp]
\centerline{\epsfxsize=8.6cm \epsfclipon \epsffile{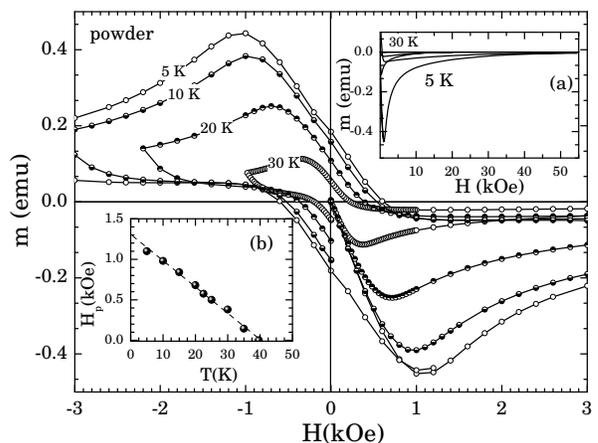}}
\caption{Variation of the magnetic moment with the magnetic field at several
temperatures (5,10, 20 and 30 K) for a powder MgB$_{2}$ sample. Inset (a)
shows half of the hysteresis loop in an extended field range for $T=5$ K and 30
K. The continuous line is a plot of the magnetic moment in the case of surface
barriers. Inset (b) shows the temperature variation of the $H_{p}$.}
\label{loopsp}%
\end{figure}

The refinement of the XRD patterns was carried out by the Fullprof Rietveld
program\cite{fullprof} using the space group $P6/mmm$. Figure \ref{rietveld}
shows the corresponding Rietveld plot of the x-ray powder diffraction
pattern. Since Mg and B occupy special positions, the only available
parameters for refinement are the unit cell constants, the occupancies and
the anisotropic temperature factors. The cell constants were found $%
a=b=3.0849(1)$\AA ~and $c=3.5213(1)$\AA . The anisotropic temperature
factors ($B_{11},B_{33},B_{12}$ in \AA $^{2}$) for Mg and B were estimated
to be ( 0.0245(1), 0.0189(1), 0.0076(1)) and (0.0164(3), 0.0(2), 0.0255(4)),
respectively. In our refinement we also included the MgO as a second phase
in order to account for a few additional small peaks. 
\begin{figure}[tbp]
\centerline{\epsfxsize=8.6cm \epsfclipon \epsffile{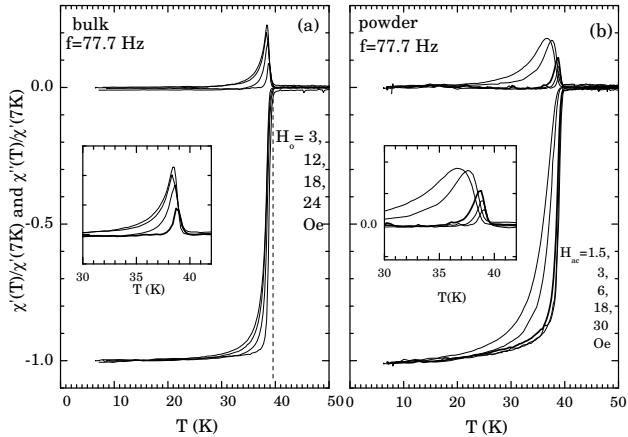}}%
\caption{Variation of the normalized real and imaginary part of the fundamental
ac-susceptibility of a bulk piece  (a) and for a powdered sample (b) of  the MgB$_{2}$ compound.
The measurements were taken at a frequency of 77.7 Hz, in zero dc-magnetic field and for
several amplitudes of the ac-field. Insets show the imaginary part of the susceptibility as a function of temperature near the $T_c$.}
\label{ac}%
\end{figure}%
This compound should
have originated from the slight excess of Mg used in the starting reaction
mixture. The estimation of the amount of this compound was 2.5 wt\%. The
occupancies for both Mg and B were found equal to one within the statistical
errors. An interesting result of the refinement is the anisotropic character
of the thermal parameters of Mg and B. We found that boron vibrates almost
in the $a-b$ plane. On the other hand the Mg has a significant component
along the $c$ axis. Finally, we must note that our structural parameters
agree with the neutron diffraction crystal data of Jorgensen et al. \cite
{0103069}.
Figure \ref{ac} shows the real and imaginary parts of the ac-susceptibility
for a bulk piece \ (nearly cylindrical in shape) \ and the same set of
measurements after the same piece was ground into fine powder. For the bulk
piece the real part, $\chi ^{\prime }$, of the ac-susceptibility for $H_{%
{\rm o}}=3$ Oe, displays a sharp drop at $T_{c}=39.7$ K. The width of the
drop in $\chi ^{\prime }$ is $\sim 0.8$ K. On the other hand the imaginary
part, $\chi ^{\prime \prime }$, of the ac-susceptibility displays a sharp
peak.
\begin{figure}[tbp]
\centerline{\epsfxsize=8.6cm \epsfclipon \epsffile{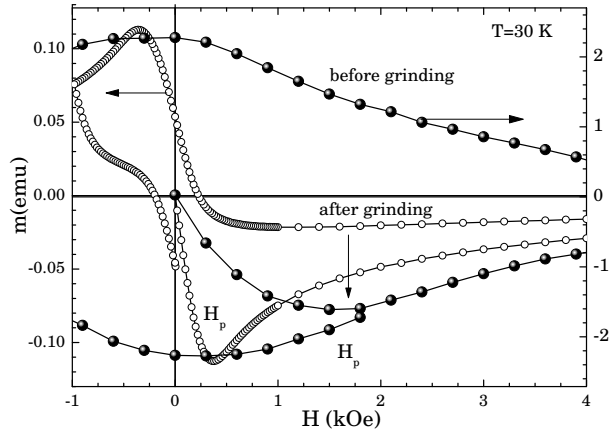}}
\caption{Variation of the magnetic moment as a function of the magnetic field
at $T=30$ K for an as prepared bulk sample and for  the same sample after grinding of
the MgB$_{2}$ compound.}
\label{loops30}%
\end{figure}%
The peak in the $\chi ^{\prime \prime }$ is located at the middle of
the drop in $\chi ^{\prime }(T)$ curve. As the amplitude of the ac-magnetic
field increases the corresponding $\chi ^{\prime }(T)$ and $\chi ^{\prime
\prime }(T)$ curves broaden and the peak of the $\chi ^{\prime \prime }(T)$
curve (or middle point of the $\chi ^{\prime }(T)$ ) is shifted to lower
temperatures. The same behavior is observed also for the powdered sample but
in this case the broadening is larger.
In order to investigate the mixed state and the magnetic irreversibility for
this material we employed isothermal magnetization measurements. Figure~\ref
{loops30} shows the magnetization loops at $T=30$ K for an as prepared bulk
piece of MgB$_{2}$ and the same measurements after grinding the bulk piece
into fine powder, in order to point out the different behavior. Figure~\ref
{loopsp} shows magnetic hysteresis loops at $T=5,10,20$ and 30 K for the
powder sample. 
The very interesting observation is the asymmetric shape of
the loops at $T=5$ K when one compares the brances for increasing and
decreasing field. In the increasing branch $-M$ drops as $\sim 1/H$, while
in the decreasing branch the magnetization is flat with values close to
zero. This behavior is different from the case where bulk pinning dominates,
giving nearly symmetric loops about the $M=0$ axis. We must note that these
features extend up to high fields (e.g. 55 kOe at $T=5$ K). The measurements
at other temperatures were similar and showed that the loop-width decreases
quickly as the temperature increases.
Wang et al.\cite{0103181} report a value for $\kappa =\lambda /\xi \approx
38 $ for MgB$_{2}$. This large value suggests that the surface barriers may
play an important role in MgB$_{2}$, even when the barriers are suppressed
by surface defects. It is instructive to compare our data with the
theoretical predictions concerning the influence of the surface barriers on
the phase diagram of MgB$_{2}$ compound. According to the theoretical
suggestions of Clem\cite{clem74} and Burlachkov et al.\cite{burlachkov94},
for the case of {\it weak bulk pinning} surface barriers may play a crucial role and determine
the first field of flux penetration as well as and the irreversibility line.
Flux penetrates through the surface by the creation of a critical nucleus
consisting of one or several vortex loops.

\begin{figure}[tbp]
\centerline{\epsfxsize=8.6cm \epsfclipon \epsffile{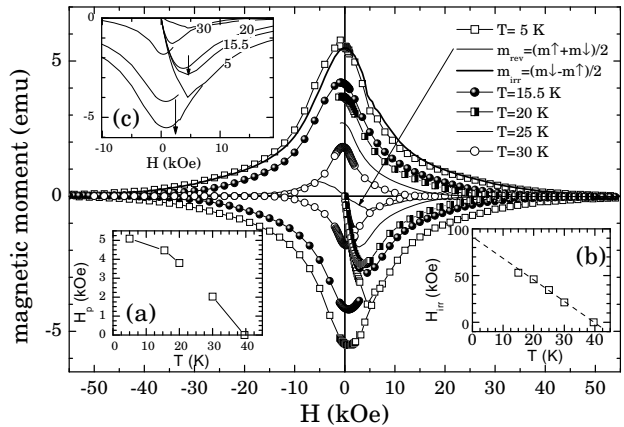}}%
\caption{Variation of the magnetic moment as a function of magnetic field at
$T=5,15.5, 20,25$ and $30$ K for the bulk MgB$_{2}$ sample. Shown are also the
irreversible and reversible magnetization as a function of magnetic field at
$T=5$ K. Insets (a-c) show the variations of $H_{fp}$, $H_{irr}$ and
 the detail of the hysteresis loops near $H=0$. respectively}
\label{loopsb}%
\end{figure}

\begin{figure}[tbp]
\centerline{\epsfxsize=8.6cm \epsfclipon \epsffile{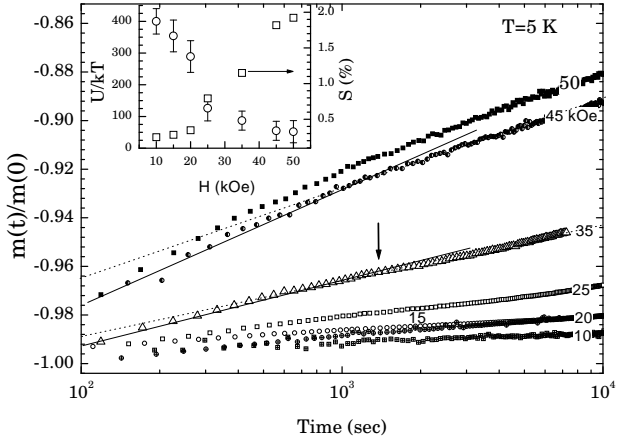}}%
\caption{Semi-logarithmic plot of the ratio $m(t)/|m(0)|$ vs time
(relaxation of magnetization) for the bulk MgB$_2$ sample
at $T=5$ K and for $H=10,15,20,25,35,45$ and $50$ kOe. Inset shows the
variation of pinning potential (open circles) and relaxation rate (squares)
 deduced from relaxation measurements with the magnetic field.}
\label{relaxation}%
\end{figure}

In the case of the powder sample we observed asymmetric loops. The
descending branch is nearly horizontal. In addition the location
of the peak of the magnetization loops can not be explained by a
model where only the reversible and irreversible magnetization are
taken into consideration. Consequently, the surface barriers must
influence the magnetic properties in the powder sample. We fitted
the peak field $H_{p}$ with the corresponding formula
$H_{p}=H_{c}(T)$ predicted from the surface barrier model ignoring
the thermal activation over the surface barrier. The inset (b) of
Fig.~\ref{loopsp} shows the variation of $H_{p}$ with temperature
extracted from magnetization measurements of the powder sample.
The variation of the $H_{p}$ with temperature is nearly
linear. According to the theoretical prediction for $H_{c}$, it varies as $%
H_{c}=\Phi _{{\rm o}}/4\pi \xi \lambda $. If we suppose a temperature
variation for $\xi $ and $\lambda $ like $(1-(T/T_{c}))^{-1/2}$ we expect
that $H_{p}=[\Phi _{{\rm o}}/4\pi \xi _{{\rm o}}\lambda _{{\rm o}%
}](1-T/T_{c})$. Namely, a linear temperature variation which is exactly what
we observe.
Despite the nice agreement of our experimental data with the concept of edge
barriers for the case of the powder samples, we can not neglect a small
contribution coming from the bulk pinning mechanism. As Brandt\cite{brandt99}
pointed out, the contribution of bulk
pinning inflates the loops nearly symmetrically about the pin-free loop. The
width $\Delta m(H=0)=m\uparrow (H=0) -m\downarrow (H=0)$ of the loop at zero
field is related to the degree of pinning, exhibiting higher values for
stronger pinning. In our case we observe that the width $\Delta m(H=0)$
increases for lower temperatures (see Fig. \ref{loopsp}).
For the bulk samples we observed symmetric magnetization loops which means
that the bulk pinning controls mainly the entry and exit of the magnetic
flux. Figure~\ref{loopsb} shows the magnetization loops for the bulk sample
at $T=5, 15.5, 20, 25$ K and 30 K. Shown are also the quantities $m_{{\rm irr}%
}=[m(\downarrow )-m(\uparrow )]/2$ and $m_{{\rm rev}}=[m(\downarrow
)+m(\uparrow )]/2$ at $T=5$ K, which represent the variation of the critical
current density and the reversible moment, respectively, as a function of
the magnetic field. Insets (a) and (b) of Fig.~\ref{loopsb} show the
temperature variation of the peak field, $H_{p}$ occuring in the virgin
magnetization loops and the irreversibility line, $H_{irr}(T)$ respectively.
$H_{p}$-curve is a measure of the temperature variation of the critical
current. $H_{irr}(T)$ line deduced from our measurements for the bulk
sample, agrees very well with those measured from other groups\cite
{finnemore01,0102436,0102353,0102517,0103302}, (extrapolates to about $\sim
80$ kOe at $T=0$ K). It seems that the $H_{irr}(T)$ curve represents a
transition of the vortex matter and not a line which depends on the pinning
strength.
The existence of the peak in the hysteresis loops at small $H$, is a
manifestation of the $B$ dependence of $J_{c}.$\cite
{brandt98,shantsev99,mcdonald96,chen89} If the critical current follows the
equation $J_{c}=J_{{\rm co}}/(1+B/|B_{{\rm o}}|)$ (Kim's model) for any
choice of the parameter $B_{{\rm o}}$, the peak is always located at
positive $H$ on the ascending branch of the loop\cite{brandt98,shantsev99}
as we observe in our measurements (see inset (c) of Fig.~\ref{loopsb} ).
In order to understand better the physical origin underlying the
magnetization loops we employed relaxation measurements of the irreversible
magnetization of the bulk sample. Specifically, we performed relaxation
measurements for several fields at $T=5$ K. In all the relaxation
measurements the sample was first cooled in zero field to the desired
temperature, and then the magnetic field was raised to the desired $H_{i}$
with a ramp rate ${\cal R}=\dot{H}_{\text{o}}\approx 100\;\text{Oe}/\text{s}$%
. After the field was stabilized to $H_{i}$, the relaxation of $m(t)$ was
measured within the time window $t_{i}=10^{2}\leq t\leq t_{f}\simeq 10^{4}\;%
\text{s}$. From the normalized relaxation rate ${\cal S}=d\ln (-m(t))/d\ln t$
we calculated the pinning potential as a function of the magnetic field at
constant temperature. Figure \ref{relaxation} is a semilogarithmic plot of
the $m(t)$ variation (relaxation of magnetization) at $T=5$ K, for $10\leq
H_{i}\leq 50$ kOe. In addition, the relaxation curves show a slope change at
a certain time. That resembles a crossover from a relaxation controlled by
bulk pinning to one controlled by surface barriers. As Burlachkov\cite
{burlachkov93} pointed out {\it the initial stage of relaxation is
determined by the weakest one of two sources of the irreversibility: the
bulk and the surface}. If the bulk pinning dominates over the surface
barrier we would expect that the initial stage is actually the surface
relaxation, where the magnetization in the surface ($M_{s}$) decreases at
approximately constant $J$. When $M_{s}=M_{{\rm eq}}$ ($M_{{\rm eq}}$ is the
equilibrium magnetization) the slope in $dM/d\ln t$ changes (decreases) and
the relaxation continues owing to the bulk mechanism. The inverse relaxation
rate, which in the framework of the interpolation formula\cite{blatter96} is
equal to ${\cal S}^{-1}=U_{c}/k_{{\rm B}}T+\mu \ln (t/t_{0})$ at small time
intervals, can give an estimation of $U_{c}/k_{{\rm B}}T$. The ${\cal S}^{-1}
$ vs $\ln t$ curves at $t=4\times 10^3$ s (not shown) are nearly
constant and decrease slightly as
the corresponding magnetic field increases. This means that $U/kT$ decreases
monotonically as the field increases. In the inset of Fig. \ref{relaxation}
plotted are the estimated values of $U/kT$ (at $t=4000$ s) for the magnetic
field where relaxation was measured.
The key question is why the as prepared bulk sample with an appreciate
porosity displays bulk pinning, while the one in the powder form does not?
One can explain this behavior with the following arguments. The crystallites
(we suppose that after thoroughly grinding we produce single crystal
particles) do not have imperfections or disorder capable to pin the flux
lines. The Ginzburg Landau coherence length for MgB$_{2}$ is $\xi \sim 40$%
\AA\ meaning that only defects of such size can pin effectively the flux
lines. Defects of such large size are difficult to be found inside the
volume of the crystallites.
On the other hand, magnetic measurements of the bulk as prepared sample,
show an enhanced critical current, indicating a substantials bulk pinning.
It seems that defects in the crystalline boundaries are able to pin the
vortices and may be the reason for the strong coupling between the grains.
Similar results with ours have been reported by Takano et al. \cite{0102167}
and Kim et al. \cite{0102338}. In these works symmetric magnetization loops
have been observed for high temperature (1000$^{{\rm o}}$C) high pressure
sintered samples. They observed asymmetric magnetization loops for the
powder and the low temperature sintered sample. Although we do not pressed
the sample it seems that the 910$^{{\rm o}}$C in the final step of the
reaction process produce the necessary pinning. Our results are also
in agreement with the conclusions of Larbalestier et al.\cite{larbalestier01},
that MbB$_2$ is not compromized by weak-link problems.

In conclusion, we presented a modified preparation method of MgB$_{2}$
compound. The magnetization loops of the powder samples are controlled by
surface barriers. In the bulk samples bulk pinning dominates rather than
surface barriers.

\acknowledgements{ This work was supported from the Greek Secretariat for
Research and Technology through the PENED program (99ED186) and Demoerevna program (program No 642).}

\end{multicols}


\begin{references}
\bibitem{akimitsu}  J.~Akimitsu, Syposium on Transition Metal Oxides.
Sendai, January 10, 2001.

\bibitem{nagamatsu}  J. Nagamatsu, N. Nakagawa, T. Muranaka,Y. Zenitani, J.
Akimitsu Nature {\bf 410}, 63 (2001).













\bibitem{0103181}  Y. Wang T. Plackowski and A. Junod, Physica C {\bf 335}, 179 (2001). 

\bibitem{Lima01}  O. F. de Lima {\it et al.}, Phys. Rev. Lett. {\bf 86},5974 (2001).
\bibitem{0105271}  M. Xu, H. Kitazawa, Y. Takano, J. Ye, K. Nishida, H. Abe, A. Matsushita and G. Kido,, cond-mat/0105271 (2001).
\bibitem{0105545}  S. Lee, H. Mori, T Masui, Yu. Eltsev, A. Yamamoto and S. Tajima, cond-mat/0105545 (2001).
\bibitem{Simon01}  F. Simon {\it et al.}, Phys. Rev. Lett. {\bf 87}, 047002 (2001).
\bibitem{0106577}  S. L. Bud'ko, V. G. Kogan and P. C. Canfield, cond-mat/0106577 (2001).
\bibitem{Patnaik}  S. Patnaik {\it et al.}, Supercond. Sci. Technol. {\bf 14}, 315 (2001).
\bibitem{0107511}  G. Papavassiliou, M. Pissas, M. Fardis, M. Karayanni and C. Christides, cond-mat/0107511 (2001).



\bibitem{0102167}  Y. Takano, H. Takeya, H. Fujii , H. Kumakura, T. Hatano
and K. Togano, Appl. Phys. Lett. {\bf 78}, 2914 (2001).

\bibitem{budko} S. L. Bud'ko, C. Petrovic, G. Lapertot, C. E. Cunningham, P. C. Canfield,
M-H. Jung and A. H. Lacerda,Phys. Rev. B {\bf 63}, 220503R (2001).

\bibitem{bugoslavsky01a} Y. Bugoslavsky, L. F. Cohen, G. K. Perkins, M. Polichetti,
T. J. Tate, R. Gwilliam, A. D. Caplin, London, {\bf 411},561, (2001). Y. Bugoslavsky, G. K. Perkins, X. Qi, L. F. Cohen and A. D. Caplin
Nature, {\bf 410}, 563 (2001).

\bibitem{cunningham01} C.E Cunningham, C. Petrovic, G. Lapertot, S. L. Bud'ko, F. Laabs,
W. Straszheim, D. K. Finnemore, P. C. Canfield, Physica C {\bf 353}, 4 (2001).

\bibitem{eom01} C. B. Eom, et al. Nature London, {\bf 411}, 558 (2001).

\bibitem{glowacki01} B. Glowacki, M. Majoros, M. Vickers, J. E. Evetts, Y. Shi and I. McDouga,
Supercond. Sci. Technol. {\bf 14},193 (2001).

\bibitem{mavoori01} S. Jin, H. Mavoori, C. Bower and R. B. van Dover, Nature London, {\bf 411}, 463 (2001).


\bibitem{kambara101} M. Kambara1, N. Hari Babu, E. S. Sadki, J. R. Cooper, H. Minami, D. A. Cardwell, A. M. Campbell and I. H. Inoue,
Supercond. Sci. Technol. {\bf 14} L5, (2001).

\bibitem{joshi01} A. G. Joshi, C. G.S. Pillai, P. Raj, S. K. Malik, Solid State Comm. {\bf 118}, 445 (2001).

\bibitem{0102338}  Mun-Seog Kim, C. U. Jung, Min-Seok Park, S. Y. Lee,
Kijoon H. P. Kim, W. N. Kang, and Sung-Ik Lee, cond-mat/0102338.

\bibitem{larbalestier01} D. C. Larbalestier et al., Nature {\bf 410}, 186 (2001).


\bibitem{fullprof}  J. Rodriguez-Carvajal, Physica B {\bf 192}, 55 (1993).

\bibitem{0103069}  J. D. Jorgensen, D. G. Hinks, S. Short, Phys. Rev. B {\bf 63}, 224522 (2001).

\bibitem{clem74}  J. R. Clem, in {\it proceeding of the 13th Conference on
Low Temperature physics} (LT 13), edited by K. D. Timmerhaus, W. J.
O'Sullivan and E. F. Hammel (Plenum, New York, 1974), Vol3, p.102.

\bibitem{burlachkov94}  L. Burlachkov, V. B. Geshkenbein, A. E. Koshelev, A.
I. Larkin, and V. M Vinokur, Phys. Rev. B {\bf 50}, 16770 (1994); L.
Burlachkov, Phys. Rev. B {\bf 47}, 8056 (1993).

\bibitem{brandt99}  E. H. Brandt, Phys. Rev. B 60, 11939 (1999).

\bibitem{finnemore01}  D. K. Finnemore, J. E. Ostenson, S. L. Bud'ko, G.
Lapertot, and P. C. Canfield, Phys. Rev. Lett. {\bf 86}, 2420 (2001).

\bibitem{0102436}  H. H. Wen, S. L. Li, Z. W. Zhao, Y. M. Ni, Z. A. Ren, G.
C. Che, H. P. Yang, Z. Y. Liu and Z. X. Zhao, Cond-mat / 0102436.

\bibitem{0102353}  Y. Bugoslavsky, G. K. Perkins, X. Qi, L. F. Cohen, and A.
D. Caplin, Nature {\bf 410}, 563 (2001).

\bibitem{0102517}  K.H. M\"{u}ller, G. Fuchs, A. Handstein, K. Nenkov, V.N.
Narozhnyi, D. Eckert, Cond-mat / 0102517.

\bibitem{0103302}  A. G. Joshi, C.G.S. Pillai, P. Raj, and S.K. Malik,
Cond-mat / 0102353.

\bibitem{brandt98}  E. H. Brandt, Phys. Rev. B {\bf 58}, 6506 (1998).

\bibitem{shantsev99}  D. V. Shantsev, et al. Phys. Rev. Lett {\bf 82}, 2947
(1999);D. V. Shantsev, Y. M. Galperin, T. H. Johansen, Phys. Rev. B, {\bf 61}%
,9699 (2000).

\bibitem{mcdonald96}  J. McDonald and J. R. Clem Phys. Rev. B, {\bf 53},
8643 (1996).

\bibitem{chen89}  D. X. Chen and R. B. Goldfard, J. Appl. Phys. {\bf 66},
2489 (1989).

\bibitem{burlachkov93}  L. Burlachkov, Phys. Rev. B {\bf 47}, 8056 (1993).

\bibitem{blatter96}  G. Blatter, V. Geshkenbein, A. Larkin, and H. Nordborg,
Phys. Rev. B {\bf 54},72 (1996); E. H. Brandt, Rep. Prog. Phys. {\bf 58,}
1465 (1995).



\end{references}
\end{document}